\documentclass{emulateapj}
\usepackage{bm}
\usepackage{color}
\usepackage{multirow}
\usepackage{graphicx}
\usepackage{longtable}
\usepackage{rotating}
\usepackage{threeparttable}
\usepackage{epsfig}

\def\OVI{O\,{\sc vi}~}
\def\OIII{[O\,{\sc iii}]~}
\def\OI{O\,{\sc i}~}

\def\HI{H\,{\sc i}~}
\def\HII{H\,{\sc ii}~}
\def\CIV{C\,{\sc iv}~}
\def\CIII]{C\,{\sc iii}]~}
\def\CII{C\,{\sc ii}~}

\def\SiIV{Si\,{\sc iv}~}
\def\SiII{Si\,{\sc ii}~}

\def\NV{N\,{\sc v}~}
\def\NII{N\,{\sc ii}~}
\def\SII{S\,{\sc ii}~}
\def\SIII{S\,{\sc iii}~}
\def\SIV{S\,{\sc iv}~}

\def\FeII{Fe\,{\sc ii}~}
\def\FeIII{Fe\,{\sc iii}~}

\def\AlII{Al\,{\sc ii}~}
\def\AlIII{Al\,{\sc iii}~}
\def\ZnII{Zn\,{\sc ii}~}
\def\MgI{Mg\,{\sc i}~}
\def\MgII{Mg\,{\sc ii}~}
\def\PV{P\,{\sc v}~}

\shorttitle{Two PDLAs with residual flux}
\shortauthors{Xie et al.}

\begin{document}
\title{An intercomparison study of two proximate damped Ly$\alpha$ systems with residual flux upon the Ly$\alpha$ absorption trough toward quasars} 
\author{Xiaoyi Xie\altaffilmark{1},
Hongyan Zhou\altaffilmark{1,2},
Xiang Pan\altaffilmark{2,1},
Peng Jiang\altaffilmark{1},
Xiheng Shi\altaffilmark{1},
Tuo Ji\altaffilmark{1},
Shaohua Zhang\altaffilmark{1},
Shengmiao Wu\altaffilmark{1},
Zhihao Zhong\altaffilmark{2,1},
}
\affil{\altaffilmark{1} SOA Key Laboratory for Polar Science, Polar Research Institute of China, 451 Jinqiao Road, Shanghai, 200136, China; zhouhongyan@pric.org.cn; xiexiaoyi@pric.org.cn;}
\affil{\altaffilmark{2} Key Laboratory for Research in Galaxies and Cosmology, Department of Astronomy, University of Sciences and Technology of China, Chinese Academy of Sciences, Hefei, Anhui 230026, China;} 

\begin{abstract}
In this paper, we present an intercomparison study of two quasars, SDSS $\rm J145618.32+340037.2$ and SDSS $\rm J215331.50-025514.1$, which have proximate damped Ly$\alpha$ systems (PDLAs) with residual flux upon the Ly$\alpha$ absorption trough. Though they both have residual flux as luminous as $10^{43}\,\rm erg\,s^{-1}$, their PDLAs are quite different in, e.g., neutral hydrogen column density, metal line absorption strength, high-ionization absorption lines as well as residual flux strength. For J$1456+3400$, the \HI column density is ${\rm log}(N_{\rm HI}/{\rm cm^{-2}}) = 20.6\pm 0.2$, with $z_{\rm abs} = 2.3138$, nearly identical to the quasar redshift ($z = 2.3142$) determined from the \OIII emission line. The metallicity of this system is typical of DLAs and there is high ionization therein, suggesting that the PDLA system is multiphase, putting it in the quasar environment. For J$2153-0255$, we measure the \HI column density to be ${\rm log}(N_{\rm HI}/{\rm cm^{-2}}) = 21.5\pm 0.1$ at $z_{\rm abs} = 3.511$, slightly redshifted with respect to the quasar ($z=3.490$) measured from C\,{\sc iii}]. The metallicity of this system is quite low and there is a lack of significant high-ionization absorption lines therein, suggesting that the system is beyond the quasar host galaxy. The residual flux is wide ($\sim \rm 1000\,km\,s^{-1}$) in J1456, with a significance of $\sim 8\sigma$, while also wide ($\sim \rm 1500\,km\,s^{-1}$) but with a smaller significance of $\sim 3\sigma$ in J2153. Among many explanations, we find that Ly$\alpha$ fuzz or resonant scattering can be used to explain the residual flux in the two sources while partial coverage cannot be excluded for J1456. By comparing these two cases, together with similar cases reported previously, we suggest that the strength of the residual flux is related to properties such as metallicity and high-ionization absorption lines of PDLAs. The residual flux recorded upon the PDLA absorption trough opens a window for us to see the physical conditions and processes of the quasar environment, and their profile and strength further remind us of their spatial scales.
\end{abstract}


\section{Introduction}
Galaxies are surrounded by vast reservoirs of gas that are capable of both emitting and absorbing \HI Ly$\alpha$ radiation \citep{Dijkstra2017}.
The brighter quasars can boost the fluorescent Ly$\alpha$ emission to detectable levels in some cases.
Through deep narrowband imaging or long-slit spectroscopic observation, there are quasars with extended Ly$\alpha$ emission have been mapped with sizes from several arcseconds \citep{Fynbo1999, Weidinger2005, Christensen2006, Herenz2015} to more than 50 arcseconds \citep{Cantalupo2014}.
The information carried by \HI Ly$\alpha$ radiation is a valuable probe to detect gas in intergalactic, circumgalactic, and interstellar media, and to understand the processes of galaxy formation and star formation and the feedback between a supermassive black hole and its quasar host galaxy.
However, as the brilliant light from a quasar also contributes to the Ly$\alpha$ radiation, it is difficult to detect light that is not extended or is too faint compared to the quasar.

Alternatively, one can use indirect methods to detect the faint but significant light that is buried by quasar light.
An ingenious way is to use damped Ly$\alpha$ systems (DLAs) as a coronagraph to shield the brilliant quasar light and reveal the faint light from the outer regions such as the star formation process in the host galaxy, cooling radiation in the galaxy haloes, scattering or fluorescence of inflow/outflow gas and so on.
The column density of neutral hydrogen for DLAs is the largest among \HI absorbers, greater than $2\times10^{20} \rm cm^{-2}$, which will produce the damped absorption wings while removing the total flux in the central part of the prominent Ly$\alpha$ absorption trough \citep{Wolfe2005}.
Thus, the light from the outer part of a quasar can be recorded as residual flux in this small wavelength range.
In fact, this concept has been verified in several DLAs \citep[e.g.,][]{Moller1998, Hennawi2009, Fynbo2010, Zafar2011, Kulkarni2012, Noterdaeme2012a, Cai2014, Fathivavsari2015, Jiang2016, North2017, Pan2017}.
In these cases, the physical parameters such as metallicity of DLAs, relative velocity between quasar/DLA and residual flux, and quasar luminosity vary from one system to another and play important roles in shaping the strength, profile, and velocity spread of the residual flux seen in the DLA absorption trough.
In fact, this reveals that the physical conditions behind this phenomenon are driven by different mechanisms and origins.

While the residual flux in the absorption trough of intervening DLAs is likely related to the star formation in the DLA galaxies \citep{Hunstead1990, Pettini1990, Kulkarni2012, Noterdaeme2012a}, the residual flux in the absorption trough of proximate DLAs (PDLAs) has varies explanations, but is mostly related to the conditions or processes in the quasar environment, e.g., emission from Ly$\alpha$ blob/fuzz \citep{Hennawi2009, Zafar2011, North2017}, partial coverage \citep{Finley2013, Fathivavsari2017}, or emission from quasar outflow gas \citep{Jiang2016}.

The variety of causes is understandable because the proximate DLAs are closer to the background quasars, less than 3000 $\rm km\,s^{-1}$ in the sense of relative velocity, and are usually considered differently from intervening DLAs due to the potential ionization conditions in the quasar neighborhood \citep{Prochaska2008, Ellison2010}.
Some of them are associated with the quasar environment and  tend to have time variability, supersolar metallicity, higher ionization parameter, partial coverage of the continuum source, and complex absorption profiles, etc \citep{Moller1998,Hamann1999, Leibundgut1999,Fox2008,Ellison2010}.
However, the division in relative velocity is not an absolute criterion because of the possibility of quasar-ejected intrinsic systems with large relative velocity or intervening systems appearing at low relative velocity \citep{Fox2008}.
In fact, there is a small fraction of PDLAs with absorption redshift greater than the emission redshift, which are then interpreted as resulting from clouds that have a peculiar inward velocity that exceeds the mean outward motion of the system of clouds \citep{Williams1970}.
\cite{Weymann1977} further invoked two separate mechanisms: intervening clouds bound in a cluster of galaxies where the quasar is embedded (for systems with an absolute relative velocity not exceeding about 3000 $\rm km\,s^{-1}$) and gravitational acceleration of material ejected by the quasar (for systems with a relative velocity of at least 3000 $\rm km\,s^{-1}$). 
For this study, the explicit knowledge of the PDLA is important in inferring the nature of the residual flux.
To know what is recorded upon the PDLA trough, one may collect the thread by examining the physical conditions of the PDLA and the manifestation of the residual flux.
 
Utill now, the number of DLAs (including PDLAs) with detection of conspicuous residual flux has been very limited compared to the tens of thousands of DLAs cataloged \citep{Noterdaeme2012b}. 
For most DLAs, the residual flux is negligible --- the average upper limit is less than $3.0\times 10^{-18} \rm erg\,s^{-1}\,cm^{-2}$ in the composite spectrum as studied by \cite{Rahmani2010}.
Thus, the rarity of this phenomenon makes it valuable because the unique configuration of quasar, DLA, and residual flux emitter allows us to study the physical processes of the outer part of the quasar environment, which generally cannot be seen under the dominant quasar light. 

In this paper, we study two quasars, SDSS $\rm J145618.32+340037.2$ and SDSS $\rm J215331.50-025514.1$, as an intercomparison to derive more clues about this issue.
Both of them are radio-quiet and have proximate DLAs with significant residual flux, but they differ in strength/profile as well as in properties of PDLAs.
We will start by introducing the data we have in Section 2, and measure the main properties of the PDLA such as column density of neutral hydrogen and metallicity as well as the residual flux in Section 3. 
Then in Section 4, we will discuss several hypotheses and judge their possibilities given the observational constraints.
Finally in Section 5, we will give a brief summary of this paper and our conclusions.
Throughout this paper, we adopt the cosmological parameters $\Omega_{\rm{\Lambda}} = 0.7$, $\Omega_{\rm M} = 0.3$, and $h = 0.7$.
 
\section{Observation}\label{sec:data}
After the identification of Ly$\alpha$ emission in the DLA trough of the mysterious quasar SDSS $\rm J095253.83+011422.0$ \citep{Jiang2016}, we have found two other quasars, $\rm J145618.32+340037.2$ (hereafter J1456) and SDSS $\rm J215331.50-025514.1$ (hereafter J2153), from Sloan Digital Sky Survey (SDSS)/BOSS data that also have significant and broad residual Ly$\alpha$ emission in the PDLA trough. 
The SDSS/BOSS spectra of J1456 and J2153 were obtained in 2011 April and 2011 November, respectively.
For the analysis below, the most recent SDSS DR13 spectra are used \citep{Albareti2017}.
With 2'' diameter fibers, the SDSS/BOSS spectra cover a wavelength range of about 3600-10,400 $\rm\AA$, with a resolution of about 2000 \citep{Smee2013}. 
The SDSS {\it ugriz} magnitudes of J1456 are 20.59, 19.75, 19.38, 19.30, and 19.01, while for the brighter J2153, they are 24.69, 19.37, 18.42, 18.14, and 18.00, respectively.
We also search the Catalina Surveys database and match these two objects (CSS\_J145618.4+340037 and CSS\_J215331.5-025513) \citep{Drake2009}.
Over a period of about 2-3 years in the quasars' rest frames, they do not show any significant variability in the $V$ band.

For J1456, besides the SDSS data, we followed this object up using the DoubleSpec (DBSP) and TripleSpec spectrographs of the P200 telescope in the Palomar observatory through the TAP (Telescope Access Program).
The DoubleSpec spectra were obtained on 2017 March 18 and 20, and took 6$\times$15 minutes in total.
We reduce the 1D spectrum following a standard IRAF\footnote{http://iraf.noao.edu/} routine.
At a similar resolution as the SDSS spectrum, the DBSP spectrum extends the wavelength coverage at the blue end to 3400~\AA. 
The residual broad Ly$\alpha$ emission is also found in the DBSP spectrum, with negligible spectral variability.
We then add the extra $3400-3600 \rm\AA$ data of the DBSP spectrum to the SDSS spectrum for J1456 in the analysis below.
Besides SDSS and the DBSP spectrum, we observed J1456 on 2016 July 15 using the TripleSpec spectrograph in the near-infrared. 
A 1"$\times$30" slit is applied and two 300s exposures were taken with dithering.
A nearby A0V star is observed for flux calibration and telluric correction. 
The SpeXtool package \citep{Cushing2004} with standard settings is applied to obtain the 1D spectrum.
This near infrared spectrum covers a wavelength range of about 11300-24630 $\rm\AA$ with a resolution of $\sim 3000$, where H$\alpha$ and \OIII emission lines are measured to determine the redshift of the quasar. 

\section{Measurement}\label{sec:measurement} 
In this section, we will first measure the quasar redshift by fitting emission lines and then measure the properties of the PDLA such as neutral hydrogen column density, residual Ly$\alpha$ flux, and metallicity.
    
\subsection{Quasar redshift}
A precise redshift for a quasar is important for the diagnosis of the PDLA properties and their relation to the quasar.
From a close visual inspection, the redshift for J2153 is estimated to be 3.49 \citep{Paris2017}.
Our fitting to the \CIII] emission line gives a consistent quasar redshift value with the pipeline.
Then we shift the spectrum to its rest frame, which extends from 800 to 2300$\rm\AA$, after correcting for the galactic extinction using the mean Milky Way extinction curve \citep{Cardelli1989} with $E(B-V)= 0.0483$ from the dust maps of \cite{Schlafly2011}. 

\begin{figure}[htb]
\hspace*{-1cm}
\epsscale{1.4}
\plotone{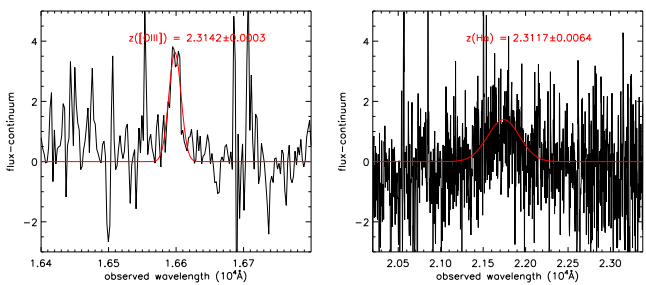}
\caption{The \OIII and H$\alpha$ emission lines in the TripleSpec spectrum of J1456. Their Gaussian fittings are plotted in red.}
\label{j1456_figure1_ir}
\end{figure}

For J1456, we measure the quasar redshift by fitting \OIII$\lambda$5007 and H$\alpha$ emission lines in the TripleSpec spectrum.
We omit the H$\beta$ and \OIII $\lambda$4959 lines since they are in a region of low signal-to-noise ratio that is heavily contaminated by telluric lines.
The H$\alpha$ is also influenced, but it is wide so we can still use a Gaussian to fit it.
Luckily, the data quality near the center of \OIII$\lambda$5007 is good, and here we recognize a narrow component.
However, it is not clear whether the line also has a broad component since the fluctuation at the outer part of \OIII is large.  
In some quasars, there is a blueshifted outflow component of [O\,{\sc iii}], but in J1456 the profile of \OIII is symmetric and thus it can be regarded as the system redshift.
In Figure~\ref{j1456_figure1_ir}, we show the \OIII and H$\alpha$ emission line regions with the local continuum subtracted and their Gaussian fitting in red.
We derive the redshift of \OIII and H$\alpha$ as $z_{\rm OIII}=2.3142\pm 0.0003$ and $z_{\rm H\alpha}=2.3117\pm 0.0064$ using their vacuum wavelengths of 5008.24 and 6564.62 $\rm\AA$, respectively.
We adopt a quasar redshift of 2.3142 from \OIII since it is narrow with high signal-to-noise ratio and no sign of outflow.
Then we shift the spectrum to its rest frame after correcting the galactic extinction with $E(B-V)= 0.0145$.

\begin{figure}[htb]
\hspace*{-0.8cm}
\epsscale{1.4}
\plotone{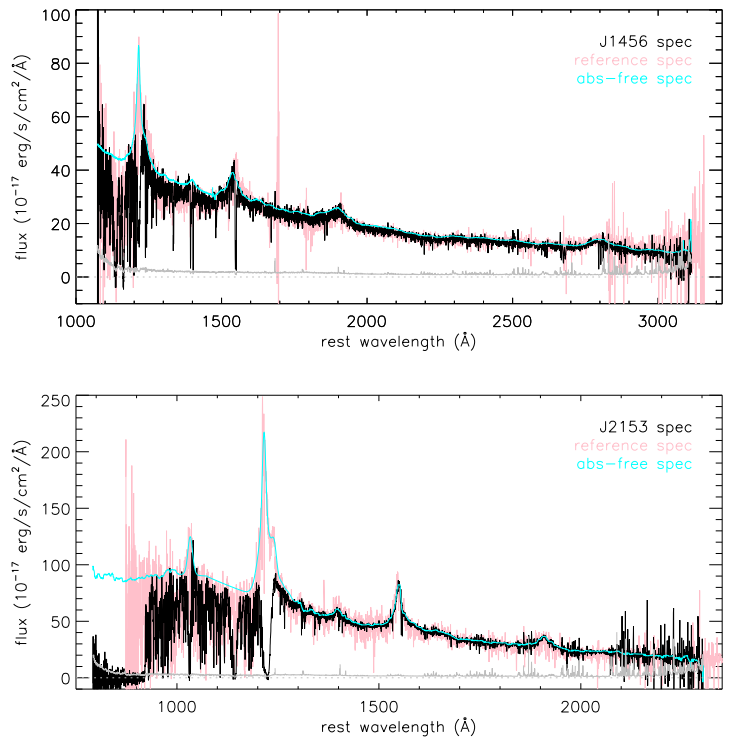} 
\caption{The observed spectrum and error in the quasar rest frame are plotted in black and dark gray. Similar spectra are plotted in pink. The estimate of the absorption-free spectrum is plotted in cyan. The upper panel is for J1456 and the lower one is for J2153.}
\label{cont_final}
\end{figure}
  
\subsection{Neutral hydrogen column density}
To derive the column density of neutral hydrogen for the absorbers in J1456 and J2153, we need to determine the absorption-free spectrum first.
In the wavelength range longer than Ly$\alpha$ emission region, the metal absorption lines are relatively narrow, so we smooth the spectra using a median filter with absorption regions masked. 
For the Ly$\alpha$ emission, as they are almost completely absorbed, we infer their profiles from quasars with similar spectra. 
To guide our eyes, we search the SDSS DR12 quasar catalog \citep{Paris2017} to find quasars with similar continuum slope and profiles of \CIV and \CIII] emission lines.
Then we visually choose the two (one) most similar spectra for J1456 (J2153), which provide a draft of the absorption-free spectrum, especially the intrinsic Ly$\alpha$ emission.
Since J2153 has a higher redshift, its high-order Lyman absorption lines and Lyman limit absorption edge are also observed.
For this short wavelength range, we modify the composite spectrum of \cite{Zheng1997} to roughly match the absorption-free level in the Ly$\beta$ region and blueward from it. 
We caution that this absorption-free level has some uncertainty due to our limited resolution and the existence of numerous Ly$\alpha$ absorbers in the sightline. 
Then we reconstruct the absorption-free spectrum by stitching the three parts together.
In Figure \ref{cont_final}, we plot the observed spectrum, a similar spectrum, and the absorption-free spectrum for J1456 and J2153.

\begin{figure}[htb]
\hspace*{-1cm}
\epsscale{1.3}
\plotone{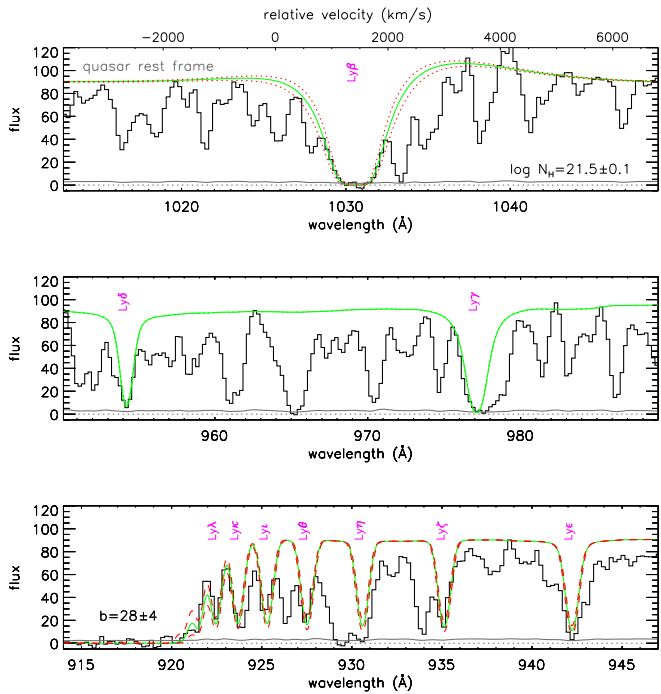}
\caption{The observed Ly$\beta$ and higher levels of \HI absorption lines up to Ly$\lambda$ for J2153. The dotted red lines show the uncertainties of $N_{\rm H}$ while the dashed red lines show the uncertainties of $b$. From high-order \HI Lyman absorption lines, we constrain the Doppler parameter to be $b=28\pm4\,\rm km\,s^{-1}$.}
\label{plot_j2153_output}
\end{figure}

In the spectrum of J1456 (J2153), we identify one (one) proximate and two (one) intervening Lyman absorption systems with associated metal lines.
A Voigt component with \HI column density $N_{\rm HI}$, Doppler parameter $b$, and redshift $z_{\rm a}$ is assigned to each Lyman absorption system to fit the identified Ly$\alpha$ absorption lines after masking the spectral regions with residual fluxes in the proximate absorption troughs.
In the fitting process, we tentatively adjust the estimation of the absorption-free spectrum to make a better fit in an iterative way.
By finding the minimum $\chi^2$, we estimate the $N_{\rm HI}$ to be $10^{20.8\pm 0.2} \rm cm^{-2}$, $10^{20.5\pm 0.2} \rm cm^{-2}$, and $10^{20.6\pm 0.2} \rm cm^{-2}$ with corresponding absorption redshifts of $z_{\rm abs} = $ 2.0954, 2.1705, and 2.3138 for J1456.
Since the spectra of J1456 do not cover Lyman series lines other than Ly$\alpha$, which is not sensitive enough to constrain $b$, we will determine $b$ from metal absorption lines later.
For J2153, the $N_{\rm HI}$ is $10^{20.6\pm 0.2} \rm cm^{-2}$ and $10^{21.5\pm 0.1} \rm cm^{-2}$ at $z_{\rm abs} = $ 3.2248 and 3.5112. 
We also detect Ly$\beta$ and high-order \HI Lyman absorption lines up to Ly$\lambda$ in the PDLA system, which are shown in Figure \ref{plot_j2153_output}.
As shown in \cite{Jiang2016}, these high-order Lyman absorption lines can be applied in constraining $b$ even at low resolution, and with theses lines we constrain the Doppler parameter to be $b=28\pm 4\,\rm km\,s^{-1}$.

\begin{figure}[htb]
\hspace*{-1cm}
\epsscale{1.4}
\plotone{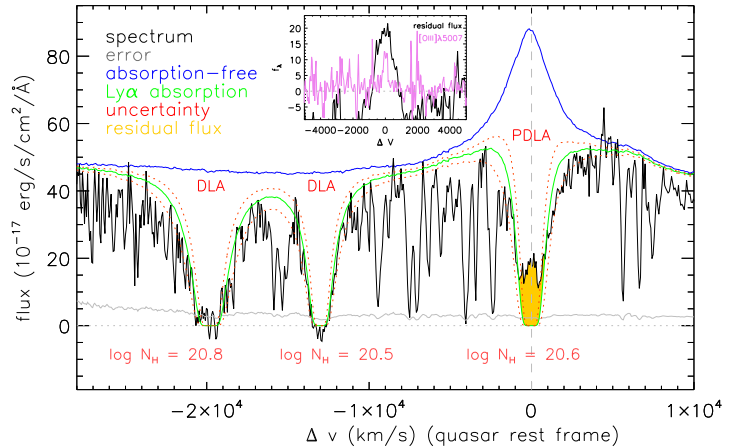}
\caption{The observed spectra and error for J1456 in the velocity space centered in the quasar rest frame are plotted in black and gray. The absorption-free spectra are shown in blue. The Ly$\alpha$ absorption and uncertainties for the DLAs are shown by green solid and red dotted lines. The residual flux of the proximate DLA is shaded in orange. In the inset, we plot the profile of the residual flux in the Ly$\alpha$ absorption trough and the \OIII$\lambda$5007 profile in violet.}
\label{j1456_3dla}
\end{figure}

\begin{figure}[htb]
\hspace*{-1.cm}
\epsscale{1.4}
\plotone{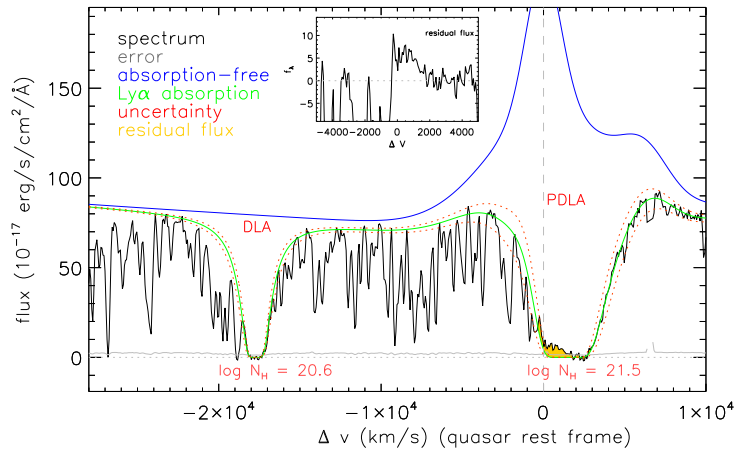}
\caption{Similar to Figure \ref{j1456_3dla}, but for J2153.}
\label{j2153_2dla}
\end{figure}

\subsection{Residual flux upon the Ly$\alpha$ absorption trough}\label{sec:residual}
We find that in the two PDLA absorption troughs there is residual flux (shaded in orange in Figure~\ref{j1456_3dla} and Figure~\ref{j2153_2dla}), in contrast to the intervening ones that have both similar \HI column density and close observed wavelength.
In the range where the observed flux is greater than in the reconstructed model, the integrated flux density of the residual flux is $f_{\rm res} = (8.54\pm 1.11)\times 10^{-16}\,\rm erg\,s^{-1}\,cm^{-2}$, equal to a luminosity of $L_{\rm res} = (3.52\pm 0.46)\times 10^{43}\,\rm erg\,s^{-1}$ for J1456.
For J2153, the corresponding values are $(4.23\pm 1.21)\times 10^{-16}\,\rm erg\,s^{-1}\,cm^{-2}$ and $(4.70\pm 1.35)\times 10^{43}\,\rm erg\,s^{-1}$.
Thus, the residual flux in J1456 has a significance of about 8$\sigma$, while it is about 4$\sigma$ in J2153.
Though the luminosity of residual flux is larger in J2153, its ratio to the continuum flux is smaller than in J1456.
In J1456, the residual flux is concentrated in the quasar/PDLA rest frame with a single strong peak, while in J2153, it has a larger relative velocity and appears in the blue edge of the PDLA absorption trough.
We further extract the residual flux from the observed spectra and use Gaussian functions to fit the profiles.
For J1456, the fitted Gaussian function has an FWHM \footnote{We subtract the instrumental broadening of 150 $\rm km\,s^{-1}$ in a quadratic way from the fitting FWHM to get the intrinsic value.} of $1082\pm 64\,\rm km\,s^{-1}$ with $z_{\rm res} = 2.3138\pm 0.0002$ ($\Delta v_{\rm res} = -35\pm 23 \,\rm km\,s^{-1}$)\footnote{$\Delta v_{\rm res} \approx c(z_{\rm res}-z_{\rm q})/(1+z_{\rm q})$ is the relative velocity between the residual flux and the quasar. Similarly $\Delta v_{\rm abs}$ is the relative velocity between the PDLA and the quasar.}.
The width of the residual flux is significantly greater than that of \OIII$\lambda$5007, which is only $356\pm 63\,\rm km\,s^{-1}$.
For J2153, the FWHM is $1554\pm 258 \,\rm km\,s^{-1}$ with $z_{\rm res} = 3.4987\pm 0.0016$ ($\Delta v_{\rm res} = 584\pm 108 \,\rm km\,s^{-1}$).
In the insets of Figure~\ref{j1456_3dla} and Figure~\ref{j2153_2dla}, we plot the profile of the residual flux in the Ly$\alpha$ absorption trough, while for J1456 we also plot the \OIII$\lambda$5007 profile in violet. 

\subsection{Metal absorption lines}
In this section, we identify metal absorption lines and assess the metallicities and dust depletion level of the two PDLAs.

\subsubsection{J1456}
\begin{figure*}[htb]
\hspace*{0cm}
\epsscale{0.9}
\plotone{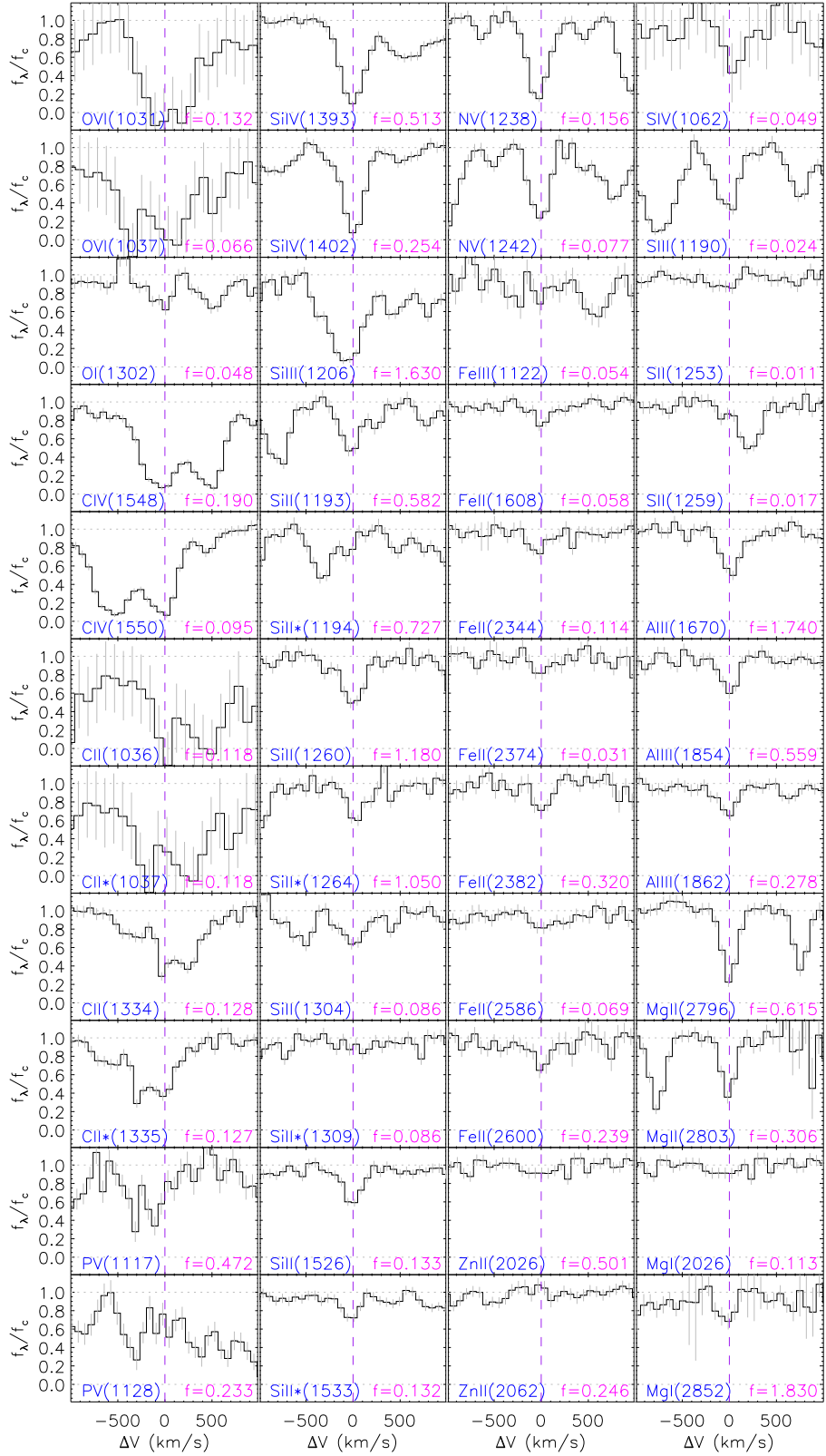}
\caption{Metal absorption lines for the SDSS spectrum of J1456 in velocity space of the absorber's rest frame. The line name and the oscillator strength are labeled in blue and magenta. The observed normalized profiles are in black and their associated measurement errors are in dark gray.}
\label{j1456_elements_row}
\end{figure*}

Metal absorption lines are abundant in the PDLA system of J1456.
Besides \CIV, \SiIV, \OI, \SiII, \CII and \SII, which are usually found in other DLAs, less common lines are also detected such as \OVI, \NV, and the metastable lines Si\,{\sc ii}* and C\,{\sc ii}*, suggesting a multiphase state of the PDLA cloud.
In Figure~\ref{j1456_elements_row}, we plot these metal absorption lines normalized by their local continuum in velocity space centered on the absorption redshift.
The observed absorption profiles are dominated by one velocity component without complex structure or this may be eliminated by the instrumental broadening.
Besides the main component, there is a weaker absorption feature at about $\Delta V = 500 \,\rm km\,s^{-1}$ in some of the lines, e.g., \NV $\lambda\lambda$1238, 1242, \CIV $\lambda\lambda$1548, 1550.
Near the \CIV emission line, there are several other weak \CIV absorption lines, revealing absorbers with small column densities in the sightline.

\begin{figure}[htb]
\hspace*{-0.6cm}
\epsscale{1.3}
\plotone{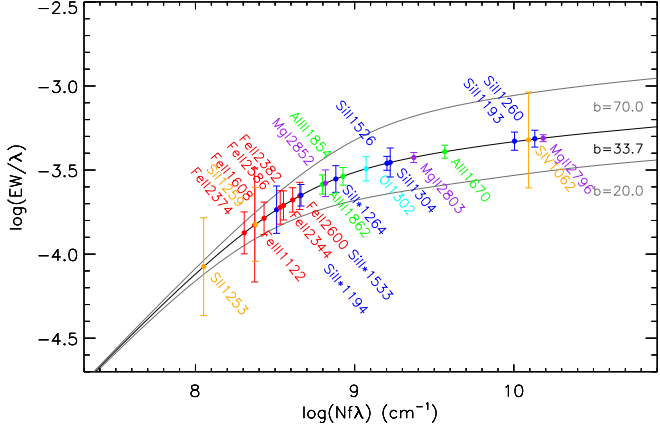}
\caption{Curve of growth for metal lines in the PDLA of J1456 with $b = 33.7\,\rm km\,s^{-1}$ in black and $b$ = 20 and 70\,$\rm km\,s^{-1}$ in dark gray. Different elements are labeled in different colors.}
\label{plot_cog1}
\end{figure}

Since the limited resolution makes the metal lines broader than they actually are and prevents a precise Voigt profile fitting, we instead adopt the curve of growth (COG) method to estimate the column density.
We omit fitting the lines that lack good quality (e.g., \OVI$\lambda$1031, \CII$\lambda\lambda$1036, 1037), are blended (e.g., \SIII$\lambda$1190 + \SiII$\lambda$1190), are below the 3$\sigma$ detection level (e.g., \ZnII$\lambda\lambda$2026, 2062, \MgI$\lambda$2026, Si\,{\sc ii}* $\lambda$1309) or hard to confirm (e.g., \PV$\lambda\lambda$1117, 1128).
For the chosen lines, we use Gaussian fitting to derive their equivalent widths (EW).
Then we constrain the $b$ parameter from the four \SiII lines and the six \FeII lines because the two groups have the same ground states.
By calculating a series of models with different $N_{\rm Si II}$, $N_{\rm Fe II}$ and $b$ values, we find that the best fitting values are $b = 33.7^{+6.3}_{-7.5}\,\rm km\,s^{-1}$ and $b = 15.9^{+3.4}_{-5.0}\,\rm km\,s^{-1}$ using \SiII and \FeII multiple lines, respectively.
The discrepancy may be explained by the fact that the \FeII multiple lines are located near the linear part, thus they are not very sensitive to $b$.
Therefore, $b = 33.7~\rm\,km\,s^{-1}$ determined from \SiII is preferred; if this is not correct, it provides a lower limit to the column density of the metal lines.
In Figure \ref{plot_cog1}, we plot the COG for the lines that are not heavily saturated using $b = 33.7~\rm km\,s^{-1}$, and also curves for $b$ = 20 and 70\,$\rm km\,s^{-1}$ for comparison.
We derive the column density from the $b = 33.7~\rm km\,s^{-1}$ curve, and estimate the errors from the uncertainties of both $b$ and EW.
In table \ref{table1}, we list the line transition, vacuum wavelength, oscillator strength $f$, equivalent width EW, and column density for these lines of J1456, where atomic data are taken from \cite{Morton2003}.

Generally, Zn provides a relatively robust estimate of metallicity since it is hardly depleted onto dust grains and is dominated by the singly ionized state in DLAs.
But in J1456, \ZnII$\lambda$2026 is contaminated with \MgI$\lambda$2026, and the absorption may be dominated by the latter because \MgI$\lambda$2852 is strong while \ZnII$\lambda$2062 absorption is not significant.
A 3$\sigma$ upper limit on Zn equivalent width is estimated from \ZnII$\lambda$2062, which is $0.21\rm\AA$ and places \ZnII$\lambda$2062 in the linear part of the COG, where absorption lines are not heavily saturated and the uncertainties are mainly induced by the measurement of EW.
This value converts to a 3$\sigma$ upper limit estimation of $N_{\rm ZnII} = 1.13\times 10^{20}\times {\rm EW}/(\lambda^2\times f) = 10^{13.36} \rm cm^{-2}$ and [Zn/H] = 0.22.
For sulfur, both \SII$\lambda$1253 and \SII$\lambda$1259 are located in the linear part of the COG, and they give a column density of about $10^{14.83\pm 0.16} \rm cm^{-2}$.
Using the solar abundance of S and Si \citep{Asplund2009}, we estimate [S/H] and [Si/H] from \SII and Si\,{\sc ii}, which are $[{\rm S/H}] = {\rm log}(N_{\rm S}/N_{\rm H})-{\rm log}(N_{\rm S}/N_{\rm H})_{\sun} = -0.87\pm 0.16$ and $[{\rm Si/H}] = -1.09\pm 0.27$ if no ionization correction is applied.
Similarly, the iron abundance estimated from multiple \FeII lines is about $[{\rm Fe/H}] = -1.96\pm 0.09$, i.e., 1/100 solar value.
The PDLA shows high dust depletion probed by [Fe/S] of -1.09, which indicates that significant amounts of dust must be present therein.

\begin{figure}[htb]
\hspace*{-0.6cm}
\epsscale{1.2}
\plotone{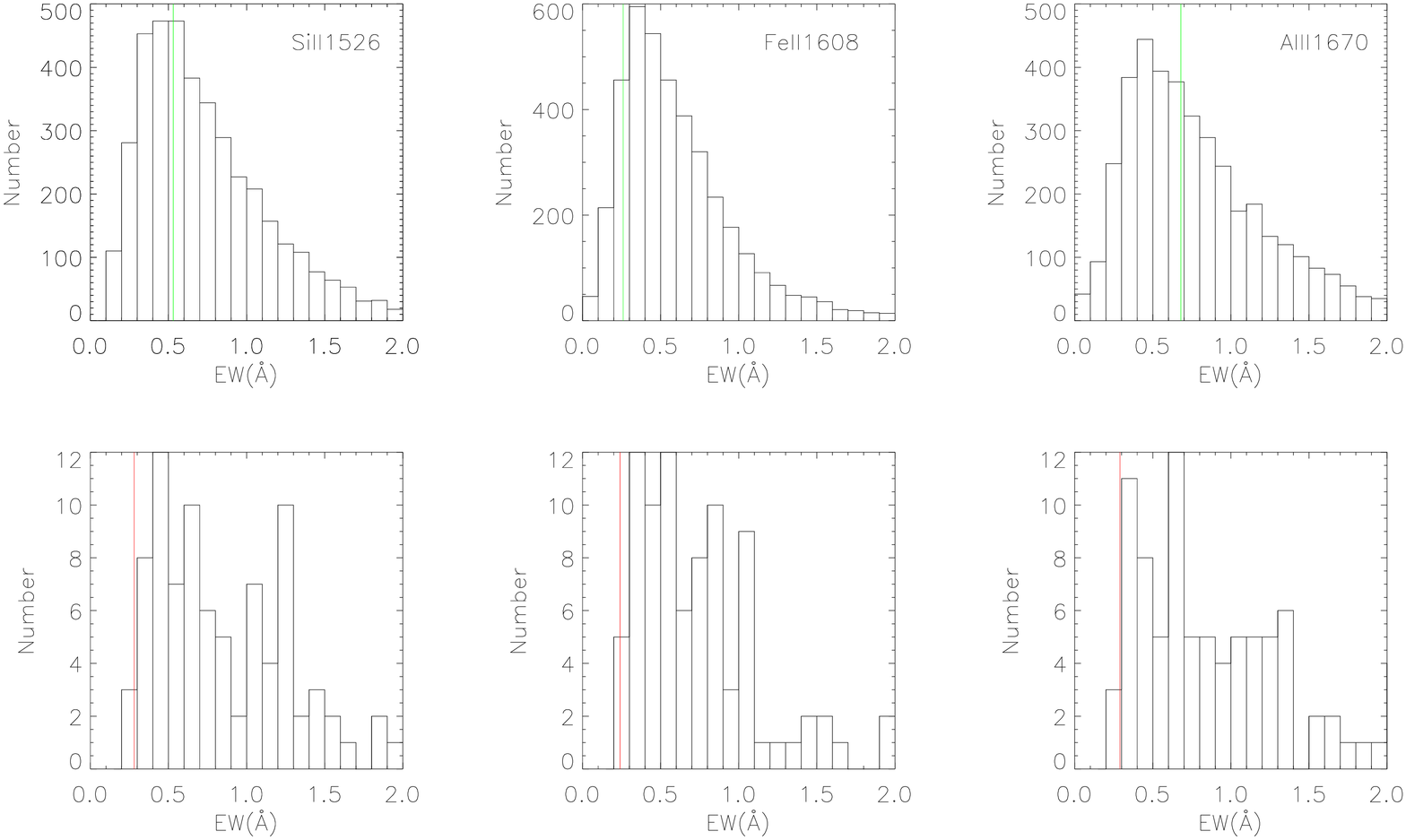}
\caption{Top panel: distribution of EW(\SiII$\lambda$1526), EW(\FeII$\lambda$1608), and EW(\AlII$\lambda$1670) of intervening DLAs with $N_{\rm HI}=10^{20.6\pm 0.2} \rm cm^{-2}$ and $z_{\rm a}=2.31\pm 0.30$ --- values comparable to J1456's. The vertical green line indicates the value of the PDLA in J1456. Bottom panel: similar to the top panel but for DLAs with $N_{\rm HI}=10^{21.5\pm 0.2} \rm cm^{-2}$ and $z_{\rm a}=3.51\pm 0.30$ --- values comparable to J2153's. The vertical red line indicates the value of the PDLA in J2153.}
\label{DLA_cat}
\end{figure}

In fact, the proximate DLA toward J1456 may be multiphase, and elements such as S and Si are probably influenced by the strong ionization in this case, while Fe also has large depletion, and thus the metallicity using \SII, \SiII, and \FeII may have some uncertainties.
To assess the metal enrichment level, we make a comparison of the equivalent widths of \SiII$\lambda$1526, \FeII$\lambda$1608, and \AlII$\lambda$1670 with those of intervening DLAs with similar $N_{\rm HI}$ and absorption redshift compiled by \cite{Noterdaeme2012b}.
The results are shown in the upper panel of Figure \ref{DLA_cat}, where we can directly see that in the distribution of EW(\SiII$\lambda$1526) and EW(\AlII$\lambda$1670), J1456 does not deviate from the median value too much while EW(\FeII$\lambda$1608) is below the median value, which indicates that the metallicity of PDLA in J1456 is typical among DLAs but with some dust depletion.
The existence of dust can be checked by comparing the relative color of J1456 to that of quasars at similar redshift.
We find the relative color is $\Delta(g-r)=0.34$ with respect to the median color of a quasar sample centered at $z=2.314$ with bin size of 0.02 from the SDSS DR12 quasar catalog, which supports the idea that there is some dust reddening of J1456.

Thus we conclude that the PDLA in J1456 is not similar to the intervening DLAs because of the existence of high-ionization absorption lines, while the metallicity is typical and there is also some dust depletion in this PDLA.
 
\begin{table}
\caption{Our measurement of the metal line column density for J1456.}
\hspace*{0.3cm}
\begin{tabular}{ccccc} 
\hline  
\hline  
line  & $\lambda (\rm \AA)$ & $f$ & EW $(\rm\AA)$ & ${\rm log}(N/\rm cm^{-2})$ \\
\hline  
\OI 1302&  1302.17&  0.048&  0.42 $\pm$ 0.07& 15.28 $\pm$ 0.38\\
\SIV 1062&  1062.66&  0.049&  0.51 $\pm$ 0.33& 16.37 $\pm$ 2.33\\
\SII 1253&  1253.81&  0.011&  0.11 $\pm$ 0.07& 14.92 $\pm$ 0.41\\
\SII 1259&  1259.52&  0.017&  0.19 $\pm$ 0.10& 15.06 $\pm$ 0.34\\
\SiII 1193&  1193.29&  0.582&  0.56 $\pm$ 0.07& 15.16 $\pm$ 0.71\\
\SiII 1260&  1260.42&  1.180&  0.61 $\pm$ 0.07& 14.96 $\pm$ 0.75\\
\SiII 1304&  1304.37&  0.086&  0.46 $\pm$ 0.09& 15.17 $\pm$ 0.52\\
\SiII 1526&  1526.71&  0.133&  0.53 $\pm$ 0.05& 14.90 $\pm$ 0.39\\
Si\,{\sc ii}* 1194&  1194.50&  0.727&  0.22 $\pm$ 0.07& 13.57 $\pm$ 0.24\\
Si\,{\sc ii}* 1264&  1264.74&  1.050&  0.35 $\pm$ 0.06& 13.76 $\pm$ 0.28\\
Si\,{\sc ii}* 1533&  1533.43&  0.132&  0.34 $\pm$ 0.05& 14.36 $\pm$ 0.16\\
\FeIII 1122&  1122.52&  0.054&  0.17 $\pm$ 0.13& 14.59 $\pm$ 0.62\\
\FeII 1608&  1608.45&  0.058&  0.26 $\pm$ 0.06& 14.46 $\pm$ 0.15\\
\FeII 2344&  2344.21&  0.114&  0.46 $\pm$ 0.09& 14.13 $\pm$ 0.16\\
\FeII 2374&  2374.46&  0.031&  0.32 $\pm$ 0.09& 14.44 $\pm$ 0.18\\
\FeII 2382&  2382.76&  0.320&  0.50 $\pm$ 0.08& 13.73 $\pm$ 0.16\\
\FeII 2586&  2586.65&  0.069&  0.49 $\pm$ 0.11& 14.28 $\pm$ 0.18\\
\FeII 2600&  2600.17&  0.239&  0.58 $\pm$ 0.11& 13.86 $\pm$ 0.18\\
\AlIII 1854&  1854.72&  0.559&  0.54 $\pm$ 0.06& 13.93 $\pm$ 0.25\\
\AlIII 1862&  1862.80&  0.278&  0.48 $\pm$ 0.06& 14.10 $\pm$ 0.20\\
\AlII 1670&  1670.79&  1.740&  0.68 $\pm$ 0.06& 14.10 $\pm$ 0.56\\
\MgII 2796&  2796.35&  0.615&  1.37 $\pm$ 0.06& 14.95 $\pm$ 0.76\\
\MgII 2803&  2803.53&  0.306&  1.05 $\pm$ 0.07& 14.44 $\pm$ 0.45\\
\MgI 2852&  2852.96&  1.830&  0.75 $\pm$ 0.14& 13.10 $\pm$ 0.25\\
\hline 
\end{tabular}
\begin{tablenotes}
\footnotesize
\item[1] Note. Columns from left to right show metal line transition, vacuum wavelength, oscillator strength $f$, equivalent width EW, and column density values.
\end{tablenotes}
\label{table1}
\end{table}

\subsubsection{J2153}
\begin{figure*}[htb]
\hspace*{-0.6cm}
\epsscale{0.9}
\plotone{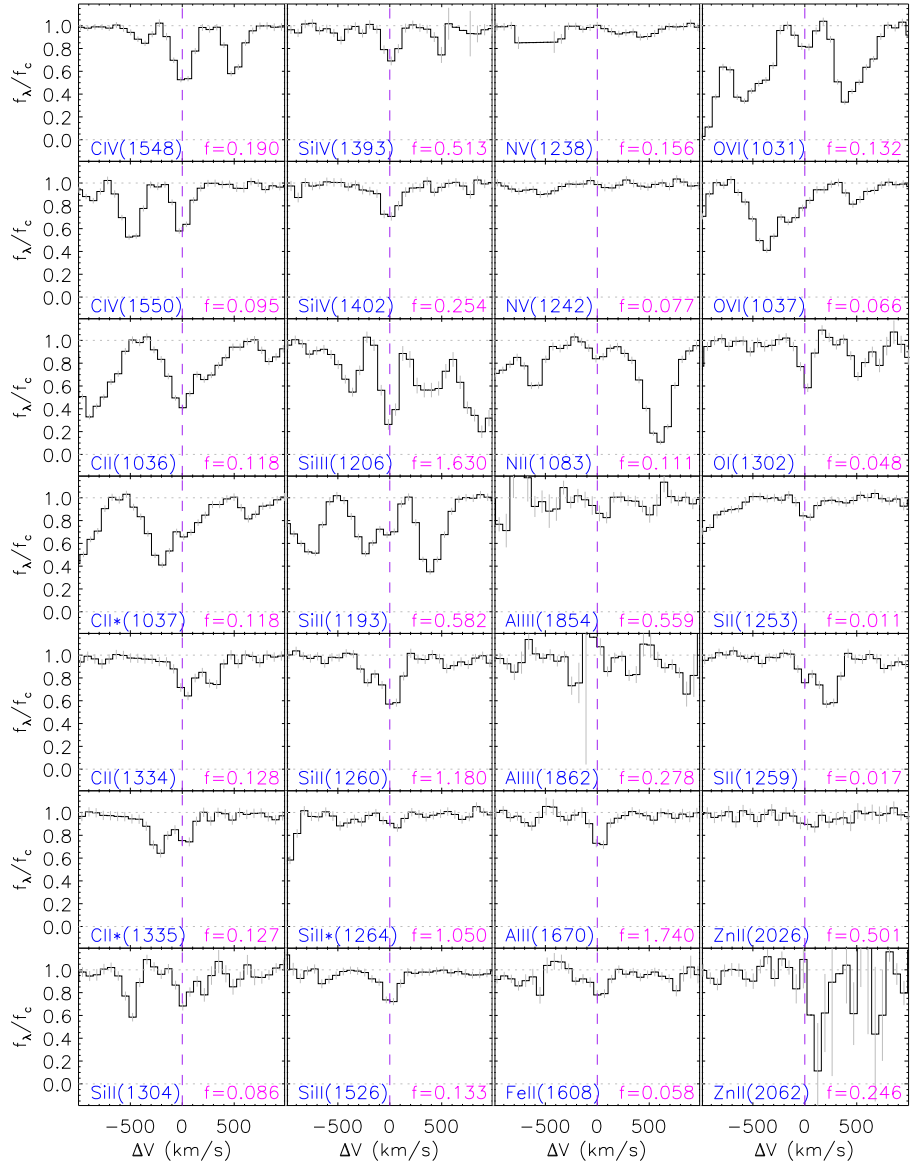}
\caption{Metal absorption lines of the SDSS spectrum of J2153 in velocity space of the absorber's rest frame. The line name and the oscillator strength are labeled in blue and magenta. The observed normalized profiles are in black and their associated measurement errors are in dark gray.}
\label{j2153_elements_row}
\end{figure*}

\begin{figure}[htb]
\hspace*{-0.6cm}
\epsscale{1.3}
\plotone{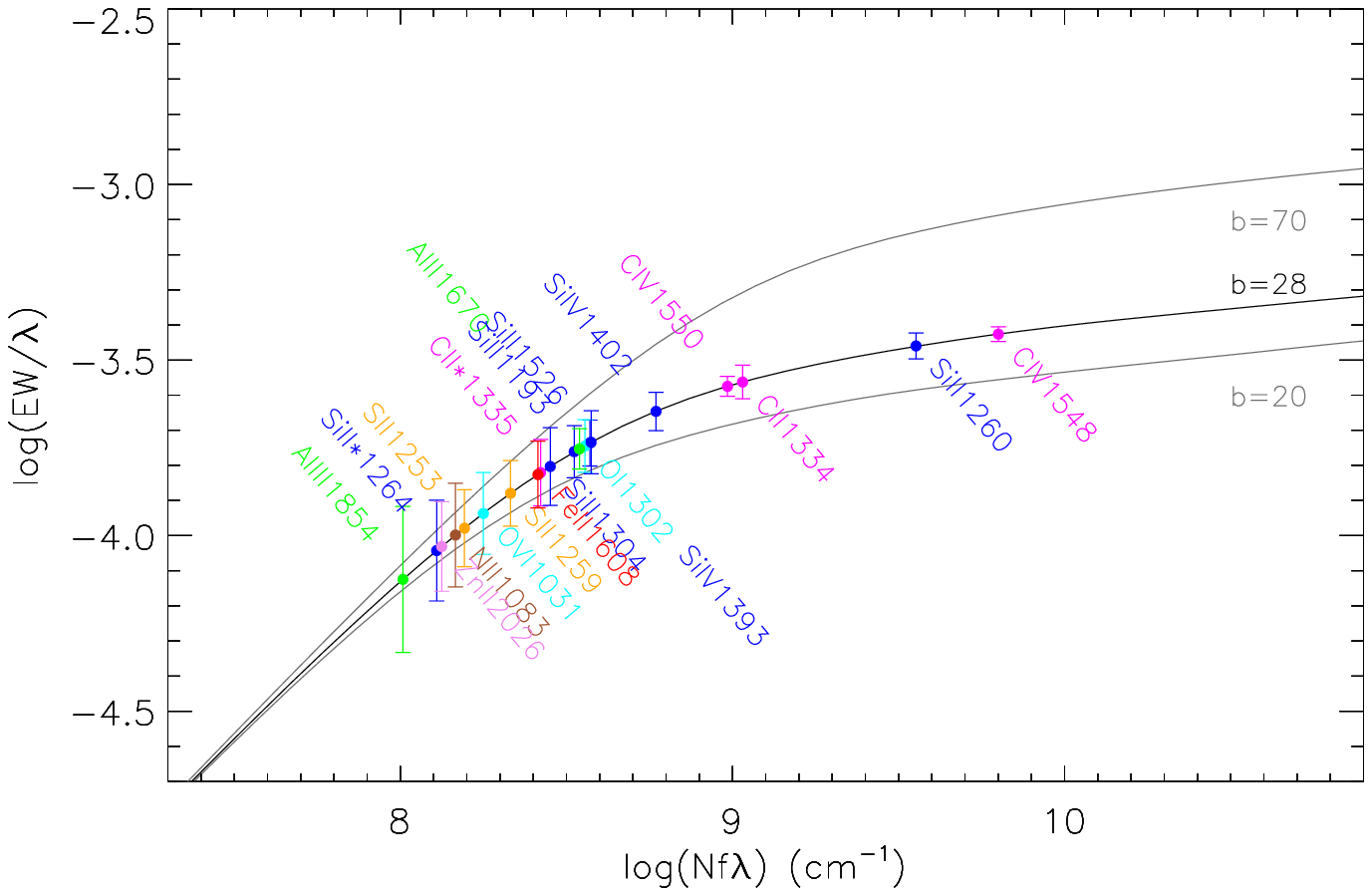}
\caption{Curve of growth of J2153 with $b = 28\,\rm km\,s^{-1}$ in black and $b = 20,70\,\rm km\,s^{-1}$ in dark grey. Different elements are labeled in different colors.}
\label{plot_cog1_2153}
\end{figure}

In the PDLA of J2153, we identify several common metal lines including \SiII$\lambda\lambda$1260, 1304, 1526, \CII$\lambda$1334, \CIV$\lambda\lambda$1548, 1550, \SiIV$\lambda\lambda$1393, 1402, \AlII$\lambda$1670, \AlIII$\lambda$1854, and \FeII$\lambda$1608.
The majority of them are low-ionization species, while high-ionization species such as \NV are consistent with non-detection.
The presence of the high-ionization lines \OVI$\lambda\lambda$1031, 1037 is uncertain since they are contaminated by the Ly$\beta$ absorption region.
We plot these metal lines normalized by their local continuum in the velocity space centered in the absorber frame in Figure \ref{j2153_elements_row}.
Like J1456, we model the absorption line with a Gaussian function separately to derive the EW value.
Then we measure the column density for the detected lines from the COG using the best fitted value $b = 28\pm 4\,\rm km\,s^{-1}$.
We list the results in Table \ref{table2} and plot the COG in Figure \ref{plot_cog1_2153}. 

The direct estimation of $[{\rm Zn/H}]$ from \ZnII$\lambda$2026 is $-0.92\pm 0.06$, which is an upper limit since the spectrum does not cover the \MgII$\lambda\lambda$2796, 2803 and \MgI$\lambda$2852 absorption lines to infer the contribution of \MgI$\lambda$2026.
For \ZnII$\lambda$2062, it has a 3$\sigma$ upper limit of equivalent width of $0.17\rm\AA$.
This is in the linear part of the COG, which converts to $N = 1.13\times 10^{20}\times {\rm EW}/(\lambda^2\times f) = 10^{13.26} \rm cm^{-2}$ and [Zn/H] $ = -0.78$.
Since the column density of \HI in the PDLA of J2153 is high, the dominant species is low-ionization absorption lines.
We estimate the sulfur and iron abundances from \SII and \FeII, and they are about $[{\rm S/H}] = -1.56\pm 0.04$ and $[{\rm Fe/H}] = -2.53\pm 0.04$.
Thus, the metallicity of this PDLA is even lower, and the dust depletion is moderate, as indicated by the relative abundance of [Fe/S] $= -0.97$.
We also make a comparison of the equivalent widths of \SiII$\lambda$1526, \FeII$\lambda$1608, and \AlII$\lambda$1670 with those of intervening DLAs with similar $N_{\rm HI}$ and absorption redshift to J2153.
The results are shown in the lower panel of Figure \ref{DLA_cat}, where we can directly see that J2153 is below the median value of the three EW distributions, which indicates that the metallicity of the PDLA in J2153 is very low.
More precise measurements of the metallicity and velocity structure of the proximate DLA absorbers need to be carried out by high-resolution spectrographs.

\begin{table}
\caption{Our measurement of the metal line column density for J2153.}
\hspace*{0.3cm}
\begin{tabular}{ccccc} 
\hline  
\hline  
line & $\lambda (\rm \AA)$ & $f$ & EW$ (\rm\AA)$ & ${\rm log}(N_{\rm line}/\rm cm^{-2})$ \\
\hline     
\CIV 1548&  1548.20&  0.190&  0.58 $\pm$ 0.03& 15.33 $\pm$ 0.46\\
\CIV 1550&  1550.78&  0.095&  0.41 $\pm$ 0.03& 14.82 $\pm$ 0.20\\
\CII 1334&  1334.53&  0.128&  0.37 $\pm$ 0.04& 14.80 $\pm$ 0.27\\
C\,{\sc ii}* 1335&  1335.70&  0.127&  0.20 $\pm$ 0.04& 14.19 $\pm$ 0.16\\
\NII 1083&  1083.99&  0.111&  0.11 $\pm$ 0.04& 14.09 $\pm$ 0.20\\
\OVI 1031&  1031.93&  0.132&  0.12 $\pm$ 0.03& 14.11 $\pm$ 0.17\\
\OI 1302&  1302.17&  0.048&  0.23 $\pm$ 0.04& 14.76 $\pm$ 0.15\\
\SII 1253&  1253.81&  0.011&  0.13 $\pm$ 0.03& 15.06 $\pm$ 0.15\\
\SII 1259&  1259.52&  0.017&  0.17 $\pm$ 0.04& 15.01 $\pm$ 0.14\\
\SiIV 1393&  1393.76&  0.513&  0.26 $\pm$ 0.04& 13.72 $\pm$ 0.14\\
\SiIV 1402&  1402.77&  0.254&  0.32 $\pm$ 0.04& 14.22 $\pm$ 0.18\\
\SiII 1193&  1193.29&  0.582&  0.21 $\pm$ 0.04& 13.68 $\pm$ 0.14\\
\SiII 1260&  1260.42&  1.180&  0.44 $\pm$ 0.04& 14.38 $\pm$ 0.43\\
\SiII 1304&  1304.37&  0.086&  0.21 $\pm$ 0.05& 14.40 $\pm$ 0.19\\
\SiII 1526&  1526.71&  0.133&  0.28 $\pm$ 0.06& 14.27 $\pm$ 0.19\\
Si\,{\sc ii}* 1264&  1264.74&  1.050&  0.11 $\pm$ 0.04& 12.99 $\pm$ 0.19\\
\FeII 1608&  1608.45&  0.058&  0.24 $\pm$ 0.05& 14.45 $\pm$ 0.16\\
\AlIII 1854&  1854.72&  0.559&  0.14 $\pm$ 0.07& 12.99 $\pm$ 0.27\\
\AlII 1670&  1670.79&  1.740&  0.29 $\pm$ 0.04& 13.08 $\pm$ 0.12\\
\ZnII 2026&  2026.14&  0.501&  0.19 $\pm$ 0.06& 13.12 $\pm$ 0.17\\
\hline  
\end{tabular}
\label{table2}
\end{table}

\section{Discussion}\label{sec:discussion}
In our study, we find two quasars, J1456 and J2153, which exhibit significant residual Ly$\alpha$ flux in the PDLA absorption trough.
In this section, we will discuss the location of the PDLA and the emitting candidate/mechanism of the residual Ly$\alpha$ flux. 
 
\subsection{Location of the PDLAs} 
For J1456, we find that the metallicity of the PDLA is subsolar, and there is the presence of \NV$\lambda\lambda$1238, 1240, which is an indicator of the metal-rich environment near quasars and hard ionization conditions different from those of intervening DLAs \citep{Hamann1999, Fox2008}.
In addition, the redshift of the quasar, the PDLA, and the residual flux emitter are almost identical.
These all suggest that the PDLA of J1456 is influenced by radiation emitted by the quasar.
We infer that the PDLA has a large probability in the quasar host galaxy, but we still need a high-resolution spectrum to confirm it.

In the case of J2153, the high redshift ($z=3.5112$) and the low metallicity of the PDLA make it an ideal lab to study the metal enrichment at high redshift.
We also notice that for this PDLA, the absorber's redshift is redshifted about 1400 $\rm km\,s^{-1}$ relative to the quasar.
This value may be an upper limit since the \CIII] emission line we use to determine the redshift has a systematic shift of about several hundred $\rm km\,s^{-1}$ \citep{Vanden2001}.
Since the absorber must lie in front of the quasar (or we will not see the absorption features), the redshifted motion can only be explained by infalling motion or just peculiar motion that deviates from the Hubble flow.
If this absorber is infalling toward the quasar, then it is probably part of the cooling flow from the circumgalactic medium surrounding the quasar, which supplies the gas for the quasar to grow.
Alternatively, if the negative relative velocity is caused by the peculiar motion, then the absorber is more likely to be a nearby galaxy in the quasar neighborhood which is a sign of an overdense region.
However, it is not possible to determine which of the two deductions is to be preferred in this study.
Considering that the metallicity is low and there is also a lack of high-ionization lines such as \NV$\lambda\lambda$1238, 1242, we infer that the PDLA is at least beyond the host galaxy, and we leave a precise conjecture of its nature to a future study.
 
\subsection{Origin of the residual Ly$\alpha$ flux}
From the analysis in the last section, we infer that the PDLA of J1456 is located in the quasar host galaxy while the PDLA of J2153 is matter infalling to the quasar or is located in a nearby galaxy.
Their residual fluxes upon the Ly$\alpha$ absorption trough are different in strength, being about 20\% of the continuum flux in J1456 and less than 10\% in J2153.
Their profiles are also different, being centrally peaked in J1456 and skewed to the blue edge in J2153.
Besides the Ly$\alpha$ absorption trough, we do not see any significant residual flux at wavelengths below the Lyman limit for these two sources, as implied by photometry from the Galaxy Evolution Explorer within the 3$\sigma$ level in the far-UV and near-UV bands \citep{Paris2017}.
Thus the residual flux existing only in the Ly$\alpha$ absorption trough is not produced by a continuum source.
In the following, we will discuss several other hypertheses that will produce the Ly$\alpha$ emission of the residual flux.

\subsubsection{By star formation process in the host galaxy}
It is reasonable to doubt that the residual flux may be caused by the star formation process in the \HII regions in either the host galaxy or in the PDLA.
If we assume that the photoionized Ly$\alpha$ photons have an escape rate of 100\% and no extinction, we can estimate the lower limit of the corresponding H$\alpha$ flux and luminosity.
Applying the case B recombination assumption (${\rm Ly\alpha}/{\rm H\alpha}=8.3$) to the observed Ly$\alpha$ residual flux, we estimate that $f$(H$\alpha$) and $L$(H$\alpha$) are greater than $1.0\times 10^{-16} \rm erg\,s^{-1}\,cm^{-2}$ and $4.2\times 10^{42} \rm erg\,s^{-1}$ for J1456 and greater than $5.1\times 10^{-17} \rm erg\,s^{-1}\,cm^{-2}$ and $5.7\times 10^{42} \rm erg\,s^{-1}$ for J2153.
Using an empirical relationship between star formation rate (SFR) and luminosity of H$\alpha$, i.e., SFR = $L(\rm H\alpha)/(1.26\times 10^{41})$ \citep{Kennicutt1998}, we estimate that the SFRs required to account for the residual flux are about $33.7\pm 4.4\rm\,M_\sun\,yr^{-1}$ and $45.2\pm 1.3\rm\,M_\sun\,yr^{-1}$.
If there is dust attenuation, the actual SFR would be larger than the estimated ones.
These values are a bit larger than the mean SFR of the host galaxy, which is about $9\rm\,M_{\sun}\,yr^{-1}$ \citep{Ho2005,Cai2014}, while a star forming galaxy can easily achieve several hundred $\rm\,M_{\sun}\,yr^{-1}$.
However, it seems that the velocity extent of the residual flux is too large to be explained by the star formation process alone.
To confirm or reject this hypothesis, we need to map the Ly$\alpha$ emitter from narrowband imaging and take a slit spectrum therein to see whether the observed H$\alpha$ flux is consistent with the inferred value.
 
\subsubsection{By partial coverage}
Recently, \cite{Fathivavsari2017} used XSHOOTER to study a quasar that has a PDLA system but with no apparent absorption in the Ly$\alpha$ region, and they ascribed the reason to the partial coverage of the broad emission line region.
In their case, the \CIV doublets are saturated in a flat-bottom shape but float on a non-zero level, which clearly indicates the partial coverage.
For J1456, we tentatively fit the wing of the \CIV$\lambda$1548 absorption with a Voigt profile convolved with instrumental broadening and find that the minimum flux should reach the zero-level in the center.
However, in the observed \CIV$\lambda$1548 absorption trough there is about 5\% residual flux, which may be a consequence of partial coverage.
Though this fraction is too small to reach the observed residual flux level in the Ly$\alpha$ absorption trough after multiplication by the absorption-free spectrum, we still cannot rule out this possibility since the \CIV emission region is closer to the center of the quasar, and then the covering factor would be greater than that of Ly$\alpha$.
We then compare the width of the residual flux to the width of the \OIII emission line, and find that the former is at least three times the latter, which means that the residual flux is not an uncovered narrow Ly$\alpha$ emission line, but could be a leaked portion of the broad emission line.
Unluckily, the resolution of the IR spectrum in the H$\alpha$ region is not high enough to produce a component fitting, which would be useful for finding a component with similar width to the residual flux.
We should test this possibility further with high-resolution spectroscopy, especially in the Ly$\alpha$, \CIV, and H$\alpha$ regions.
For J2153, due to its higher redshift, we examine the residual flux in the Ly$\beta$ region and find that it is consistent with zero.
Thus we find that partial coverage is not suitable for J2153.

\subsubsection{By radiation from Ly$\alpha$ fuzz}
In the early stage of galaxy formation, there is supposed to be a phase where galaxies are surrounded by a spatially extended distribution of infalling cold gas \citep{Haiman2001}.
When a central quasar is formed, its strong radiation will lead to the excitation of Ly$\alpha$ emission and form a Ly$\alpha$ fuzz.
In the case of radio-quiet quasars, this Ly$\alpha$ fuzz is even fainter and less extended, and much harder to detect.
Observationally, such giant Ly$\alpha$ fuzz has only been detected in very rare cases, as compiled by \cite{Herenz2015}.
According to the calculation of \cite{Haiman2001}, the typical angular diameter for this Ly$\alpha$ fuzz is a few arcseconds and its surface brightness is about $10^{-18}$ to $10^{-16} \rm erg\,s^{-1}\,cm^{-2}\,arcsec^{-2}$.
If we take the SDSS fiber size of 2 arcsec, then the surface brightness of J1456 and J2153 will be greater than $(2.71\pm 0.35)\times 10^{-16} \rm erg\,s^{-1}\,cm^{-2}\,arcsec^{-2}$ and $(1.50\pm 0.43)\times 10^{-16} \rm erg\,s^{-1}\,cm^{-2}\,arcsec^{-2}$.
Also the observed velocity extent for the residual flux is consistent with the typical range of Ly$\alpha$ fuzz, which extends to $1000-1500\,\rm km\,s^{-1}$ \citep{Hennawi2009}.
Since both of these two quasars are radio-quiet, we consider the Ly$\alpha$ residual flux in the DLA trough to be close to that predicted by the Ly$\alpha$ fuzz, especially for J2153 where the residual flux is redshifted with respect to the quasar.

\subsubsection{By resonant scattering of outer gas}
Since Ly$\alpha$ is a resonance line that has a large transition probability, when a Ly$\alpha$ photon is emitted, it shortly recombines to its prior state.
Thus, it is also likely to be a strong scattering line.
In this resonant scattering process, the Ly$\alpha$ photons absorbed by the PDLA are re-emitted therein, with no loss in energy but only a change in direction.
This could explain the phenomenon that the residual flux is found only in the Ly$\alpha$ trough of our data.
Observation of polarization can be a good method to discriminate the resonant scattering line since the polarization of scattered Ly$\alpha$ photons depends strongly on whether Ly$\alpha$ is resonantly scattered or not \citep{Dijkstra2008}.
For example, \cite{Hayes2011} detected polarized Ly$\alpha$ from the blob LAB1, which helped them to confirm that these Ly$\alpha$ photons are produced in the galaxies hosted within the nebula.
If the residual flux is caused by the resonant scattering, we expect to check this possibility from polarization observations.

\subsection{Analogs of J1456 and J2153}
The PDLA of J1456 is characterized by typical metallicity and the presence of \NV absorption lines, whereas the residual flux is characterized by a small velocity shift, is centerally peaked, and has a large significance.
From the previous reporting of the residual flux detected in the PDLA absorption trough, we find other sources sharing these similarities, e.g., the PDLAs in SDSS J1240+1455 \citep{Hennawi2009} and SDSS J1047+3423 \citep{Ma2017, Ma2018}.
\cite{Hennawi2009} proposed that the residual flux in their source is due to the fluorescent recombination radiation intrinsic to the quasar's host and similar to the extended Ly$\alpha$ fuzz detected around many active galactic nuclei.
For SDSS J1047+3423, \cite{Ma2017, Ma2018} measured the MMT+Keck high-resolution data, where the flat bottoms of the \CIV and \MgII profiles are consistent with a zero-level, which excludes a partial coverage being the cause, and they suggest that the cause of the residual flux can be interpreted as being associated with star formation or Ly$\alpha$ photons scattered from the quasar.

The PDLA of J2153 is characterized by low metallicity and the absence of \NV absorption lines, whereas the residual flux is weak compared to the continuum and closer to the quasar redshift than the PDLA redshift.
Similar residual flux profiles have been reported in Q0151+048A \citep{Moller1998, Fynbo1999, Zafar2011}, which has a redshifted PDLA and residual flux in the blue edge of the absorption trough; the authors conclude that the flux seen in the DLA trough is from  host galaxy, from either stellar emission or dust reflection. 
Another similar case is Q2059-360 \citep{Leibundgut1999, North2017}, which also has a redshifted PDLA, but the residual flux is in the red edge of the absorption trough.
According to the authors, this emission line feature is also probably a Ly$\alpha$ blob, reaching $\sim 120$ kpc in extent.

\section{Summary}
In this paper, we present an intercomparison study of two PDLAs with residual flux but different properties such as absorption redshift, amount of neutral hydrogen, ionization state, chemical composition, and dust depletion level.
\begin{itemize}
\item[1.]{In the PDLA at $z_{\rm abs}=2.31$ of J1456 ($z_{\rm q}=2.31$), the \HI column density is lower (${\rm log}(N_{\rm HI}/\rm cm^{-2}) = 20.6)$, the metallicity and ionization state are higher, and there is greater dust depletion, while the residual flux is stronger and significant.}
\item[2.]{In the PDLA at $z_{\rm abs}=3.51$ of J2153 ($z_{\rm q}=3.49$), the \HI column density is higher (${\rm log}(N_{\rm HI}/\rm cm^{-2}) = 21.5)$, the metallicity and ionization state are lower and there is less dust depletion, while the residual flux is weaker and less significant.}
\end{itemize}

We infer that the PDLA of J1456 is in the quasar host galaxy while the PDLA in J2153 is likely infalling matter to the quasar or in a nearby galaxy.
From the absorption profiles, we find that the residual flux is not likely to be produced by partial coverage for J2153, but we need a high-resolution spectrum to check this for J1456.
We also check the star formation origin of this residual flux and find that the SFRs needed are about 30 $\rm M_{\sun}\,yr^{-1}$ and 45 $\rm M_{\sun}\,yr^{-1}$, respectively.
The residual flux has a velocity extent about 1000 $\rm km\,s^{-1}$ in J1456, which is broader than the \OIII emission.
For J2153, this value is even higher, about 1500 $\rm km\,s^{-1}$.
For these reasons, we suggest that the residual flux is not likely to be caused by the star formation process.
Instead, we prefer that the residual flux is due to resonant scattering from gas in the quasar host galaxy in the case of J1456 and to Ly$\alpha$ fuzz from the outer region surrounding the quasar in the case of J2153, though we need further evidence to confirm this. 
Pieces of observational evidence and interpretations analyzed above suggest that: (1) residual flux overlaid on the Ly$\alpha$ absorption trough of the PDLAs is helpful in recording and deducing the physical processes in quasars; (2) possible links exist between the profile/strength of the residual flux and the PDLA's properties; (3) the residual flux that is different in profile and strength in the proximate DLA absorption trough may trace gas on different spatial scales in the nearby environment of the quasar.
Though confidence is low in drawing a general conclusion from only two object, it will be interesting to compare more PDLAs with different properties such as absorption redshift, amount of H\,{\sc i}, ionization state, metallicity, and dust depletion to see whether there is a trend in the profile/strength of residual flux with properties of PDLAs.
Further observations such as high-resolution spectroscopy and narrowband imaging will be useful for conducting a precise measurement of metal absorption lines, mapping the residual flux emitter, and predicting the scattering level of the Ly$\alpha$ residual flux with observations of polarization.

\acknowledgments 
We acknowledge the anonymous referee for helpful comments that clarified important aspects of the manuscript. X.X.Y. thanks Prof. Zhengyi Shao and Prof. Shiyin Shen in Shanghai Astronomical Observatory for careful reading and helpful suggestions. T. Ji is supported by NSFC (grant No. 11503022) and CAS Key Laboratory of Astronomical Optics and Technology (grant No. CAS-KLAOT-KF201706). This research uses data obtained through the Telescope Access Program (TAP), which has been funded by the National Astronomical Observatories of China, the Chinese Academy of Sciences, and the Special Fund for Astronomy from the Ministry of Finance. Observations obtained with the Hale Telescope at Palomar Observatory were obtained as part of an agreement between the National Astronomical Observatories, Chinese Academy of Sciences, and the California Institute of Technology.

Funding for SDSS-III has been provided by the Alfred P. Sloan Foundation, the Participating Institutions, the National Science Foundation, and the U.S. Department of Energy Office of Science. The SDSS-III web site is http://www.sdss3.org/.
SDSS-III is managed by the Astrophysical Research Consortium for the Participating Institutions of the SDSS-III Collaboration including the University of Arizona, the Brazilian Participation Group, Brookhaven National Laboratory, Carnegie Mellon University, University of Florida, the French Participation Group, the German Participation Group, Harvard University, the Instituto de Astrofisica de Canarias, the Michigan State/Notre Dame/JINA Participation Group, Johns Hopkins University, Lawrence Berkeley National Laboratory, Max Planck Institute for Astrophysics, Max Planck Institute for Extraterrestrial Physics, New Mexico State University, New York University, Ohio State University, Pennsylvania State University, University of Portsmouth, Princeton University, the Spanish Participation Group, University of Tokyo, University of Utah, Vanderbilt University, University of Virginia, University of Washington, and Yale University. 

The CSS survey is funded by the National Aeronautics and Space Administration under Grant No. NNG05GF22G issued through the Science Mission Directorate Near-Earth Objects Observations Program. The CRTS survey is supported by the U.S.~National Science Foundation under grants AST-0909182 and AST-1313422.

\end{document}